\documentclass[apl,showpacs,amsmath,amssymb,twocolumn,final]{revtex4-1}

\usepackage[dvips]{graphicx,color}% Include figure files
\usepackage{dcolumn}% Align table columns on decimal point
\usepackage{bm}% bold math
\usepackage{ulem}
\usepackage{soul}

\usepackage{graphicx,epstopdf}
\usepackage{dcolumn}% Align table columns on decimal point
\usepackage{bm}% bold math

\begin{document}

\title{Superconducting properties of very high quality NbN thin films grown by high temperature chemical vapor deposition }

\author{D. Hazra$^{1}$$^{,2}$, N. Tsavdaris $^{3}$, S. Jebari $^{1}$$^{,2}$, A. Grimm$^{1}$$^{,2}$, F. Blanchet$^{1}$$^{,2}$, F. Mercier$^{3}$, E. Blanquet$^{3}$, C. Chapelier$^{1}$$^{,2}$, and M. Hofheinz$^{1}$$^{,2}$}

\affiliation{$^{1}$ Univ. Grenoble Alpes, INAC-PhElIQS, F-38000 Grenoble, France.}
\affiliation{$^{2}$ CEA, INAC-PhElIQS, F-38000 Grenoble, France.}
\affiliation{$^{3}$ Univ. Grenoble Alpes, SIMAP, F-38000 Grenoble, France.}

\date{\today}

\begin{abstract}

Niobium nitride (NbN) is widely used in high-frequency superconducting electronics circuits because it has one of the highest superconducting transition temperatures ($T_c$ $\sim$ 16.5 K) and largest gap among conventional superconductors. In its thin-film form, the $T_c$ of NbN is very sensitive to growth conditions and it still remains a challenge to grow NbN thin film (below 50 \,nm) with high $T_c$. Here, we report on the superconducting properties of NbN thin films grown by high-temperature chemical vapor deposition (HTCVD). Transport measurements reveal significantly lower disorder than previously reported, characterized by a Ioffe-Regel ($k_F$$\ell$) parameter of $\sim$ 14. Accordingly we observe $T_c$ $\sim$ 17.06\,K (point of 50\,\% of normal state resistance), the highest value reported so far for films of thickness below 50\,nm, indicating that HTCVD could be particularly useful for growing high quality NbN thin films.

\end{abstract}

\maketitle

Niobium nitride (NbN)thin films --- thanks to their high $T_c\sim 16.5\,\mathrm{K}$, superconducting energy gap $\Delta\sim 2.5\,\mathrm{meV}$, and upper critical field $B_{c2}\sim 40\,\mathrm{T}$ --- have been the subject of intense research for the last few decades, both on application and fundamental grounds. The combination of high $T_c$ and small coherence length ($\xi(0) \sim 5\,\mathrm{nm}$) allows one to fabricate very thin NbN films with reasonably high $T_c$, which is essential for, e.g, Superconducting Single Photon Detectors (see e.g.\cite{Delacour-APL, G-JMO}). NbN thin films are used as hot electron bolometers and superconducting radio frequency cavities. NbN has higher kinetic inductance to other S-wave superconductors\cite{Anthony}, which this helps fabricating superconducting micro wave resonators with high characteristic impedance and microwave kinetic inductance detectors. On the fundamental level, the effects of disorder on superconducting and normal state properties have been studied in NbN thin films \cite{Mondal-PRL,Chand-PRB,Noat}. Nano-wires, made from NbN thin films, have demonstrated thermal and quantum phase slips \cite{Delacour-Nanolett}--- a phenomenon of great interest in understanding one-dimensional superconductivity. Further, the large superconducting energy gap of NbN can be explored in designing circuit Quantum Electrodynamics experiments in the THz frequency range.

Thus, there has been a growing demand of high quality NbN thin films. Reactive DC magnetron sputtering from an Nb target in an argon and nitrogen atmosphere is most commonly used to deposit NbN on various substrates\cite{Gavaler, Wang-JAP, Chockalingam-PRB}.  The main difficulty in this process, arises from the creation of atomic level nitrogen vacancies and from the formation of non-superconducting Nb$_{2}$N and hexagonal phases. Besides, in the optimal parameter range, the high sputtering rate (typically $\sim$ 1-5 nm/sec) makes it difficult to control the thickness below 10\,nm. Some other methods, where the superconducting properties of NbN thin films were probed, include Pulsed Laser Deposition (PLD) \cite{Treece, Senapati}, Molecular Beam Epitaxy (MBE) \cite{Lin-PRB} and Atomic Layer Deposition (ALD) \cite{Mario-SST}. In this regard, deposition of superconducting NbN films by high temperature chemical vapor deposition (HTCVD) is rather rare. HTCVD, compared to most of the other methods, has certain advantages: it is cost effective, especially for large scale productions and the growth rate being tunable to low values (down to 1 nm/minute), it is easy to control the thickness of the films.

A first step to explore HTCVD as an alternative to the existing methods to produce good quality superconducting NbN films, we grow three 40-nm thick NbN films by HTCVD, and investigate their superconducting properties. The free electron density ($n$) of the films is determined from Hall measurements at room temperature. The zero temperature upper critical field $B_{c2}(0)$ and the Ginzburg-Landau coherence length $\xi(0)$ are estimated from magneto resistance data near $T_c$. The strong coupling nature of our NbN films is confirmed by superconducting energy gap ($\Delta$) measurement of a film by scanning tunnel spectroscopy at 1.35\,K. We also estimate $\lambda(0)$ from the normal state resistivity ($\rho_{xx}$) and $\Delta(0)$. Our best film has very high $k_F\ell \sim 14$ and $T_c \sim 17.06\,\mathrm{K}$, and quite low $\rho_{xx} \sim 60\,\mu\Omega\mathrm{cm}$ and $\lambda(0)\sim 175\,\mathrm{nm}$, making it very promising for many superconducting applications.

\section {Experiments and Results}

Three 40-nm-thick NbN films, labeled as S2, S3, and S4 respectively, were grown simultaneously (under the same process condition) at $1300\,^\circ\mathrm{C}$ on different sapphire substrates by HTCVD. The detailed deposition process and the structural characterizations of the films are reported elsewhere \cite{Frédéric}. Briefly, S2 and S4 were grown on ($11\bar{2}0$) and (0001) oriented sapphire (Al$_{2}$O$_{3}$) substrate, respectively; whereas, S3 was grown on  (0001) orientated Al$_{2}$O$_{3}$ with a buffer layer of 80-nn-thick aluminium nitride (AlN) buffer layer grown by HTCVD \cite{Boichot}. X-ray diffraction and HRTEM studies reveal that all the NbN-films contain face centered cubic (FCC) as primary phase, grown preferentially along the (111) orientation. The average out-of-plane lattice parameters ($a$) are listed in Table \ref{tab1}. Apart from the (111) preferential orientation, all the films also show additional orientations (mainly 200). The f factors --- a measure of the degree of preferential orientation for FCC films \cite{f-factor} --- are also listed in Table \ref{tab1}. S2, apart from containing only cubic phases, does not contain any other phase; whereas S3 and S4 also contain a very small fraction of hexagonal phases, more in S3 than in S4.

\begin{figure}\centerline{\includegraphics[width=8.5cm,angle=0]{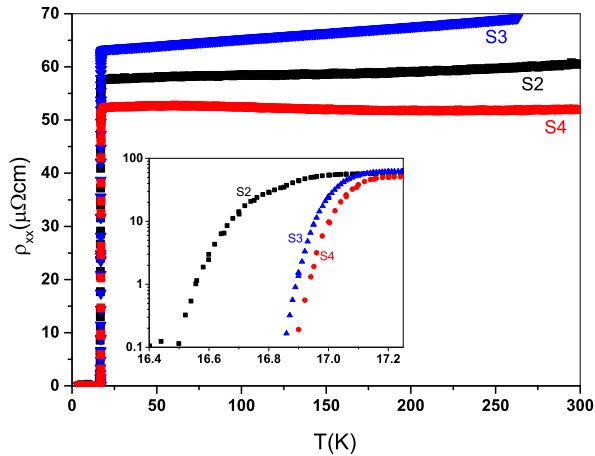}}
\caption{The temperature variation of resistivity ($\rho_{xx}$) for S2, S3, and S4. The inset shows a zoom on the superconducting transition. The $\rho_{xx}$ at 17.25 K, $T_c$, and $\Delta T_c$ are listed in Table \ref{tab1}.}
\label{fig:RT}
\end{figure}

The electrical transport measurements were performed down to 4\,K in a commercial physical property measurement system (PPMS). In Fig. \ref{fig:RT}, we plot the temperature variation of resistivity ($\rho_{xx}$) for S2, S3, and S4. In the normal state, the resistivity has very little variation in temperature. The normal state resistivities $\rho_{xx}$ of the films, defined at T = 17.25\,K, are listed in Table \ref{tab1}. The inset of Fig. \ref{fig:RT} shows the superconducting transition of resistivity (on a log scale). The $T_c$ of the films, defined as the temperatures where resistivity is half of the normal value, are listed in Table \ref{tab1}.

%In general, as Fig.\ref{fig:RT} indicates, the resistance in the normal state has a very little variation in temperature; nevertheless, for S2 and S3, the resistance at the normal state continuously decreases with temperature, whereas, for S4, the resistance shows an oscillatory behavior--- a minima occurs near 250 $K$ and a maxima near 50 $K$. The other important parameters --- the resistivity at room temperature $rho_{n}$ and the Residual Resistivity Ratio (RRR) ---  --- have also been listed in Table-\ref{tab1}. RRR is defined as $R(275K)/R(20K)$.

\begin{figure}\centerline{\includegraphics[width=8.5cm,angle=0]{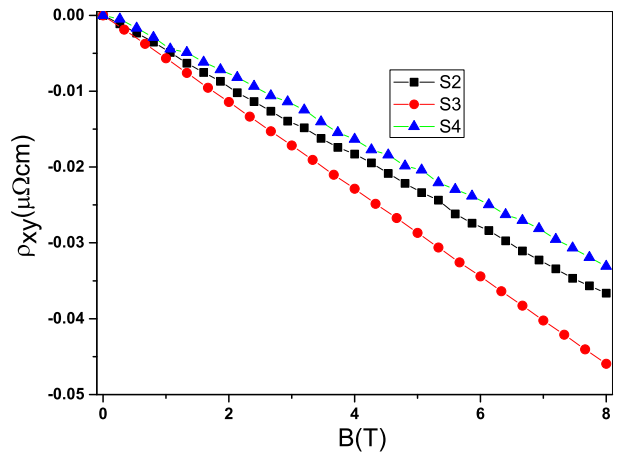}}
\caption{The hall resistivity for all three samples at room temperature. The magnetic field was scanned from 0 to 8 $T$. The electron densities, listed in Table-\ref{tab1}, are estimated from the slope ($R_H$) using the formula: $R_H = 1/ne$.}
\label{fig:Hall}
\end{figure}

In Fig.~\ref{fig:Hall}, we plot the Hall resistivity ($\rho_{xy}$) for all three samples. The electron density ($n$) was estimated from the slope, known as the Hall resistance $R_H = 1/ne$, where $e$ is the charge of an electron. The Hall measurement was performed at room temperature where electron--electron interaction is expected to be weak \cite{Khodas}, justifying the free electron approximation. $R_H$ values at room temperature for all three samples are listed in Table \ref{tab1}.  Knowing $n$ and $\rho_{xx}$, the elastic scattering time ($\tau$ ) was estimated from Drude's formula: $\rho_{xx} =m/ne^{2}\tau$, here, m is the mass of a free electron. The other important parameters --- the Fermi wave vector($k_F$), the Fermi velocity ($v_F$), the mean free path ($\ell$), and the diffusion constant ($D$) --- were estimated from the following formulae: $k_F = (3\pi^{2} n)^{1/3}$, $v_F = \hbar k_F/m$, $\ell = v_F \tau$, $D = v_F\ell/3$. In Table \ref{tab2}, we summarize these parameters together with $k_F\ell$.

% low coherence length $\xi$ $\sim$ 5 $nm$, and fast energy relaxation time

\begin{figure}\centerline{\includegraphics[width=8.5cm,angle=0]{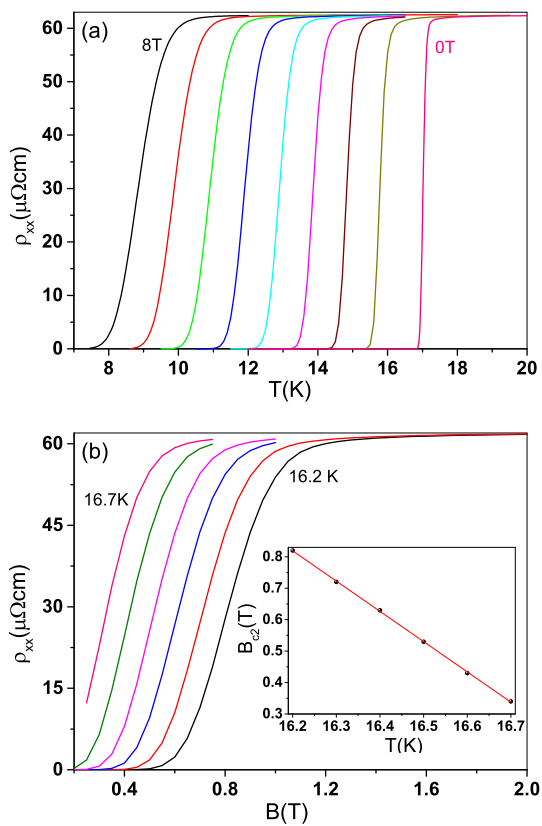}}
\caption{ (a) The variation of resistivity ($\rho_{xx}$) of S3 as a function of temperature for magnetic fields 0 to 8\,T. (b) The variation of $\rho_{xx}$ as a function of magnetic field at different temperatures, 16.2 to 16.7\,K with increments of 0.1\,K, near $T_c$. The inset shows the variation of the $B_{c2}$ as a function of temperature, with $B_{c2}$ being defined as the field where the resistance is half of the normal state resistance at 17.25\,K. The straight line fit, indicated by the solid line, gives the slope used to determine $B_{c2}(0)$ and $\xi(0)$, listed in Table \ref{tab2}.}
\label{fig:MagScan}
\end{figure}

In Fig.~\ref{fig:MagScan}, we present the magneto-resistance data for S3. Fig.~\ref{fig:MagScan}a presents the variation of $\rho_{xx}$ as a function of temperature for magnetic fields from 0 to 8\,T. The graph clearly shows that the zero temperature value of the upper critical field ($B_{c2}(0)$) is greater than 8\,T, the maximum field our set up could provide. We, therefore, extract $B_{c2}(0)$ from the following formula: $B_{c2}(0) = 0.69T_c\left|\frac{\mathrm{d}B_{c2}}{\mathrm{d}T}\right|_{T=T_c}$ \cite{Helfand}. The slope near $T_c$, $\left(\frac{\mathrm{d}B_{c2}}{\mathrm{d}T}\right)_{T=T_c}$, was extracted as shown in Fig.\ref{fig:MagScan}b. Here, we plot the variation of resistance as a function of magnetic field at different temperatures near $T_c$. The inset shows the variation of $B_{c2}$, defined as the field where the resistivity is half the normal state value of $\rho_{xx}$ (taken at 17.25 K), as a function of temperature. The slope is extracted from the straight line fit, indicated by the solid line. We also extract $\xi(0)$ from the slope, using the following formula: $\xi(0)$ = $\sqrt{\Phi_0/2\pi T_c \left|\frac{dB_{c2}}{dT}\right|_{T=T_c}} $ \cite{tinkham-book}. In Table-\ref{tab2}, we summarize both $B_{c2}(0)$ and $\xi(0)$.

%$\sqrt{\frac{\Phi_0}{2\pi T_c \left|\frac{dB_{c2}}{dT}\right|_{T=T_c} }}$

\begin{figure}\centerline{\includegraphics[width=8.5cm,angle=0]{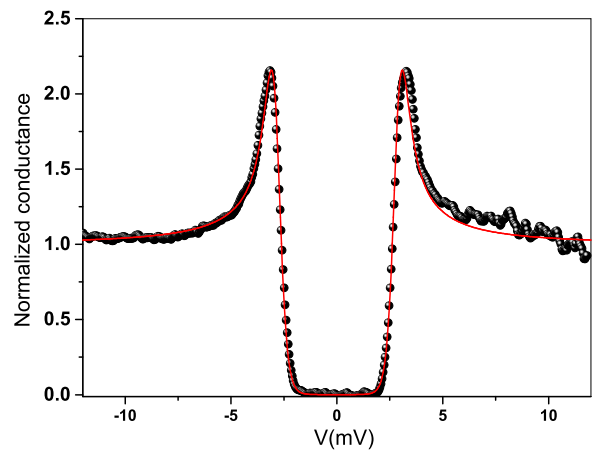}}
\caption{ Normalized tunneling conductance versus bias voltage at 1.35\,K on sample S2 (black points). The red curve is a BCS theoretical fit with a superconducting gap of 2.8\,meV.}
\label{fig:STM}
\end{figure}

In order to estimate the zero temperature superconducting energy gap ($\Delta (0)$) of our films, we have performed scanning tunneling spectroscopy (STS) measurement on S2 at $T=1.35\,\mathrm{K}$, a temperature much lower than $T_c$. $\Delta (0)$ is found to be inhomogeneous over space--- it's value ranging from 2\,meV to 2.8\,meV. The suppressed superconducting gaps might be due to local contamination or oxidation of the surface as the sample had been exposed to air over a period of more than one year prior to STS measurements . Fig.~\ref{fig:STM} shows an experimental spectrum and a fit computed with a gap of 2.8 meV according to Bardeen-Cooper-Schrieffer (BCS)\cite{BCS} theory.  This leads to a ratio 2 $\Delta$ / $k_B$$T_c$ = 4, which is slightly less than values reported in the literature \cite{Chockalingam-PRB,Noat}.

%After $T_c$, we now report on the upper critical field ($B_{c2}$) and coherence length $\xi(0)$. Since, at low temperatures, for most of the samples, $B_{c2}$ is much higher than 8 $T$ --- the maximum field our set up can provide --- we measure $B_{c2}$ only near $T_c$, and extract $B_{c2}$ and $\xi(0)$ from the following formula: $B_{c2}(0)$ = 0.69$T_c$$\left(\frac{dB_{c2}}{dT}\right)_{T=T_c}$ and $\xi(0)$ = $\sqrt{\frac{\Phi_0}{2\pi B_{c2}(0)}}$. Here, $B_{c2}(0)$ is the critical field at zero temperature and $\Phi_0$ is the flux quanta. To demonstrate, in Fig.\ref{fig:MagScan}, we present the magneto resistance data for S4. At a fixed temperature, while measuring the resistance the magnetic field was scanned from 0 to 8 $T$. The $B_{c2}$ is defined at a field where the resistance is half of the normal resistance, indicated by the dotted lines. The inset shows the variation of the $B_{c2}$ as a function of temperature near $T_c$. The straight line fit, indicated by the line, gives the slope which is used to determine the $\xi(0)$, listed in Table\ref{tab1}.

\begin{table*}
\caption{\label{tab1} Summary of the experimental results. Here, $\rho_{xx}$ is the normal state resistivity at 17.25\,K. $T_{c}$ is defined at a temperature where resistivity is half of $\rho_{xx}$.}
\begin{tabular}{|c|c|c|c|c|c|c|c|c|}
\hline
Samples &Substrate &$a$& $f$ factor & $R_{H}$& $\rho_{xx}$ & $T_{c}$ & $\Delta T_{c}$ &$\left|\frac{\mathrm{d}B_{c2}}{\mathrm{d}T}\right|_{T=T_c}$ \\
        &               &$\mathrm{\AA}$& &($10^{-3} \mu\Omega\mathrm{cm}$/T)&($\mu\Omega\mathrm{cm}$)  &  (K)  & (K)& (T/K)\\
\hline
S2       & Al$_{2}$O$_{3}$ $(11\bar{2}0)$&4.436 &0.92   & 4.6 & 57.1  & 16.80&0.55&0.96\\
S3       & AlN $(0001)$&4.434 &0.99   & 5.6 & 62.5     & 17.02&0.32&0.99\\
S4       & Al$_{2}$O$_{3}$ $(0001)$ &4.427 &0.98   & 4.1 & 51.3   & 17.06&0.32&0.96\\
\hline
\end{tabular}
\end{table*}

\begin{table*}
\caption{\label{tab2} Summary of the various parameters calculated from the experimental results of Table \ref{tab1}.}
\begin{tabular}{|c|c|c|c|c|c|c|c|c|c|c|}
\hline
Samples &  $n$ & $\tau$ & $\ell$ & D & $k_{F} \ell$ & $N_u$ &$B_{c2}(0)$&$\xi(0)$ &$\lambda(0)$\\
        &              ($10^ {29}/\mathrm{m}^{3}$)&(fS) &(nm)&($\mathrm{cm}^{2}/\mathrm{sec}$)& &($(\mathrm{eV})^{-1}$) & (T)       & (nm) & (nm)\\
\hline
S2       & 1.34&0.46&0.84&5.1&13 &0.46 & 11.1     & 6.5 & 179\\
S3       & 1.10&0.52&0.89&5.1& 13 &0.43   & 11.3      & 6.5 & 186\\
S4       & 1.49&0.46&0.87&5.5&  14 &0.48  & 11.7     & 6.4 & 169\\
\hline
\end{tabular}
\end{table*}

\section {Analysis and Discussion}

We have, so far, extensively used free electron model (FEM) to estimate various parameters like $n$ or $k_F\ell$. FEM, in general, works very well for good metals. On the other hand, unlike a good metal, our films have very little variation in resistivity as a function of temperature in the normal state (Fig.~\ref{fig:RT}). Thus, it is not obvious that the free electron model should work well. In this regard, we observe that the free electron densities $n$ of our films are significantly lower than the theoretical estimate of $n = 2.39 \times 10^{29}/\mathrm{m}^{3}$\cite{Mathur}. The other important parameter to compare, as suggested by Chockalingam et. al \cite{Chockalingam-PRB},  is the density of state (DOS) at the Fermi level. The DOS per unit volume per energy level, in the framework of free electron model, is given by: $N_V = mk_F/\hbar^{2}\pi^{2}$. The DOS per NbN unit, $N_u$, listed in Table-\ref{tab1}, is then $N_u = N_Va^{3}/4$ (each unit cell of volume $a^{3}$ is shared by 4 NbN units). Contrary to the electron density $n$, we see that the estimated DOSs are quite close to the theoretical estimate of 0.54/eV \cite{Mattheiss} or from the specific heat measurement $\approx$ 0.5/eV \cite{Geballe}.

The critical temperature $T_c \sim 17\,\mathrm{K}$ of our films is, to the best of our knowledge, the highest reported to date for NbN thin films with thickness 50\,nm or less. The $T_c$ values stated in Table-\ref{tab1} are defined as the point of 50\,\% of the normal state resistance (taken at 17.25\,K). The resistivity reaches 1\,\% of the normal state resistance at 16.60 K, 16.89 K, and 16.91 K for S2, S3, S4, respectively. These numbers are still significantly higher than previously reported numbers and show that the superconducting transition is very sharp. This, we believe, is the result of  the epitaxial nature and therefore the good crystallinity of our films, that lead to $k_F\ell$ values significantly higher than the ones reported previously (see e.g., \cite{Chockalingam-PRB, Chand}). Chand et al.\cite{Chand} reported $T_c\sim 17.0\,\mathrm{K}$ for one of their sputtered NbN films with thickness more than 50 nm. For sputtered NbN films, the enhancement of $T_c$ with thickness has been observed even above 100 nm \cite{Wang-JAP}. Chockalingam et al. \cite{Chockalingam-PRB} also observed a systematic increase in $T_c$ with $n$ or DOS at the Fermi level. We do not observe such trend. Wang et al. \cite{Wang-JAP} observed a systematic non-monotonic variation of $T_c$ with $a$; the highest $T_c$, they observed, was for $a = 4.46\,\mathrm{\AA}$.

%Here, we would like to mention that for both Chockalingam et al and Wang et al, the DC magnetron sputtered films on MgO substrate  were grown along (200) ; whereas, our HTCVD films on sapphires grow along (111). Among three films, S2 has significantly lower $T_c$ in comparison to S3 and S4. This may be due to the fact that S2, in comparison to S3 and S4, is quite less epitaxial as given by f-factor in Table-\ref{tab1}.

The upper critical field $B_{c2}$ is very similar for all of our films, but much lower than for other films with $T_c$ higher than 14 $K$ reported in the literature\cite{Chockalingam-PRB,Capone,Saraswat}. In the dirty limit --- defined by $\ell \ll \xi(0)$ --- $B_{c2}(0)$ is related to resistivity and diffusion constant via $B_{c2}(0)  = 0.69T_c4k_B/\pi e D = 0.69 T_c 4 e k_B N_V \rho_{xx}$/$\pi$ \cite{DeGennes}. The low $B_{c2}$ are, therefore, a further indication that our films have lower disorder, i.e.\ larger diffusion constant / lower resistivity. Indeed, the resistivities $\rho_{xx} \sim 60\,\mu\Omega\mathrm{cm}$ we observe are significantly lower than previously reported \cite{Chockalingam-PRB,Capone,Saraswat}. One should note, however, that the above formula yields $B_{c2}(0) \sim 2.5\,\mathrm{T}$ for our films, a significantly lower field than the value extracted from Fig.~\ref{fig:MagScan}. This discrepancy could be due to the spin-orbit interaction, which has not been taken into account in the above formula and can enhance $B_{c2}(0)$ very significantly \cite{Werthamer}.

%$B_{c2}(0)$  = 0.69$T_c$4$e$$k_B$$N(0)$$\rho_{xx}$/$\pi$ \cite{Werthamer}

%$B_{c2}(0)$  = 0.69$T_c$4$k_B$/$\pi$e$D$

%Despite quantitative disagreement, the above formula has been found to be in qualitative agreement with experiment \cite{Chockalingam-PRB}.

Finally, we estimate the zero temperature value of the magnetic penetration depth ($\lambda(0)$) from $\Delta(0)$, using the following formula \cite{Kamlapure}:
$\lambda^{-2}(0) = \pi\mu_0\Delta(0)/\hbar\rho_{xx}$. For NbN films, this formula is found to closely match experiment \cite{Kamlapure}. The calculated $\lambda(0)$s are listed in Table-\ref{tab1}. In our calculation, $\Delta(0)$ is estimated from strong coupling relation $2\Delta(0)/k_B T_c = \alpha$ --- assuming $\alpha$ = 4.00 for all three films. S4, as we see from Table-\ref{tab1}, has minimum $\lambda(0)$ among three films, which is due to its highest $T_c$ and lowest $\rho_{xx}$ values.

In summary, we have found exceptionally high critical temperatures around 17\,K and large superconducting gap of 2.8\,meV for NbN films grown by HTCVD at $1300\,^\circ\mathrm{C}$ on sapphire and AlN. We explain this high critical temperature by very large $k_F\ell$ parameters indicating low disorder. Consistent with this interpretation, we observe low resistivity ( $\sim$ $60\,\mu\Omega\mathrm{cm}$) and low upper critical field ($\sim$11\,T). Our results demonstrate that HTCVD, a particularly cost effective growth technique, is a very promising alternative to magnetron sputtering for depositing high-quality NbN thin films. A natural extension of this work will be to to further investigate the existing links between the material characteristics and the superconducting properties and to explore the limits of the HTCVD techniques for the production of high quality ultra thin films of NbN.

Acknowledgements: We acknowledge financial support from French National Research Agency / grant ANR-14-CE26-0007 - WASI, from the Grenoble Nanosciences Foundation grant JoQOLaT and from the European Research Council under the European Union's Seventh Framework Programme (FP7/2007-2013) / ERC Grant agreement No 278203 – WiQOJo, as well as fruitful discussions within the WASI project with Luca Redaelli, Eva Monroy, Val Zwiller and Jean-Michel Gérard. DH acknowledges the fruitful discussions with Mintu Mondal.
%$\lambda(0)$ = $2^{1/2}$ $\lambda_{GL}(0)$ $133^{1/3}$ $\eta_{\Delta(0)}^{-1/2}$. $\lambda_{GL}(0)$ --- the weak coupling GL penetration depth --- can be estimated from the formula : $\lambda_{GL}(0) (cm)$  = 6.24 $\times$ $10^{-6}$[$\rho_{n}(\mu \Omega cm) T_{c}(K)] ^{1/2}$. $\eta_{\Delta(0)}^{-1/2}$,  related to $\alpha$ with 3.53  $\eta_{\Delta(0)}^{-1/2}$ =

\end{document}